\documentclass[12pt, a4paper]{article} % 10pt font size (11 and 12 also possible), A4 paper (letterpaper for US letter) and two column layout (remove for one column)

\usepackage[utf8]{inputenc}
\usepackage[english]{babel} % English language hyphenation

\usepackage{url}
\usepackage[breaklinks]{hyperref}

\usepackage{biblatex}
\usepackage{color}
\usepackage[margin=1.0in]{geometry}

\usepackage{graphicx}
\usepackage{authblk}
\usepackage{csquotes}

\title{MDEmic in a use case for microscopy metadata harmonization: Facilitating FAIR principles in practical application with metadata annotation tools}

\author[1,*]{Susanne Kunis}
\author[2]{Sebastian H\"ansch}
\author[3]{Christian Schmidt}
\author[4]{Frances Wong}
\author[5]{Caterina Strambio-De-Castillia}
\author[2]{Stefanie Weidtkamp-Peters}

\affil[1]{Department of Biology/Chemistry and Centre for Cellular Nanoanalytics, University Osnabrueck, 49076 Osnabrueck, Germany}
\affil[2]{Centre for Advanced Imaging, University Duesseldorf, 40225 Duesseldorf, Germany}
\affil[3]{Bioimaging Centre, Department of Biology, University of Konstanz, 78457 Konstanz, Germany}
\affil[4]{Division of Computational Biology, Centre for Gene Regulation and Expression, University of Dundee, Dundee UK}
\affil[5]{Program in Molecular Medicine, University of Massachusetts Medical School, Worcester MA 01605, USA}

\date{} %display no date

\begin{document}
\maketitle % Print the title
\subsection*{Abstract}
While the FAIR principles are well accepted in the scientific community, the implementation of appropriate metadata editing and transfer to ensure FAIR research data in practice is significantly lagging behind. On the one hand, it strongly depends on the availability of tools that efficiently support this step in research data management. On the other hand, it depends on the available standards regarding the interpretability of metadata. Here, we introduce a tool, MDEmic, for editing metadata of microscopic imaging data in an easy and comfortable way that provides high flexibility in terms of adjustment of metadata sets. This functionality was in great demand by many researchers applying microscopic techniques. MDEmic has already become a part of the standard installation package of the image database OMERO as OMERO.mde. This database helps to organize and visualize microscopic image data and keep track of their further processing and linkage to other data sets. For this reason, many imaging core facilities provide OMERO to their users. We present a use case scenario for the tailored application of OMERO.mde to imaging data of an institutional OMERO-based Membrane Dye Database, which requires specific experimental metadata. Similar to public image data repositories like the Image Data Resource, IDR, this database facilitates image data storage including rich metadata which enables data mining and re-use, one of the major goals of the FAIR principles.\\

\begin{figure}[h]
\centering
\includegraphics[width=1.0\linewidth]{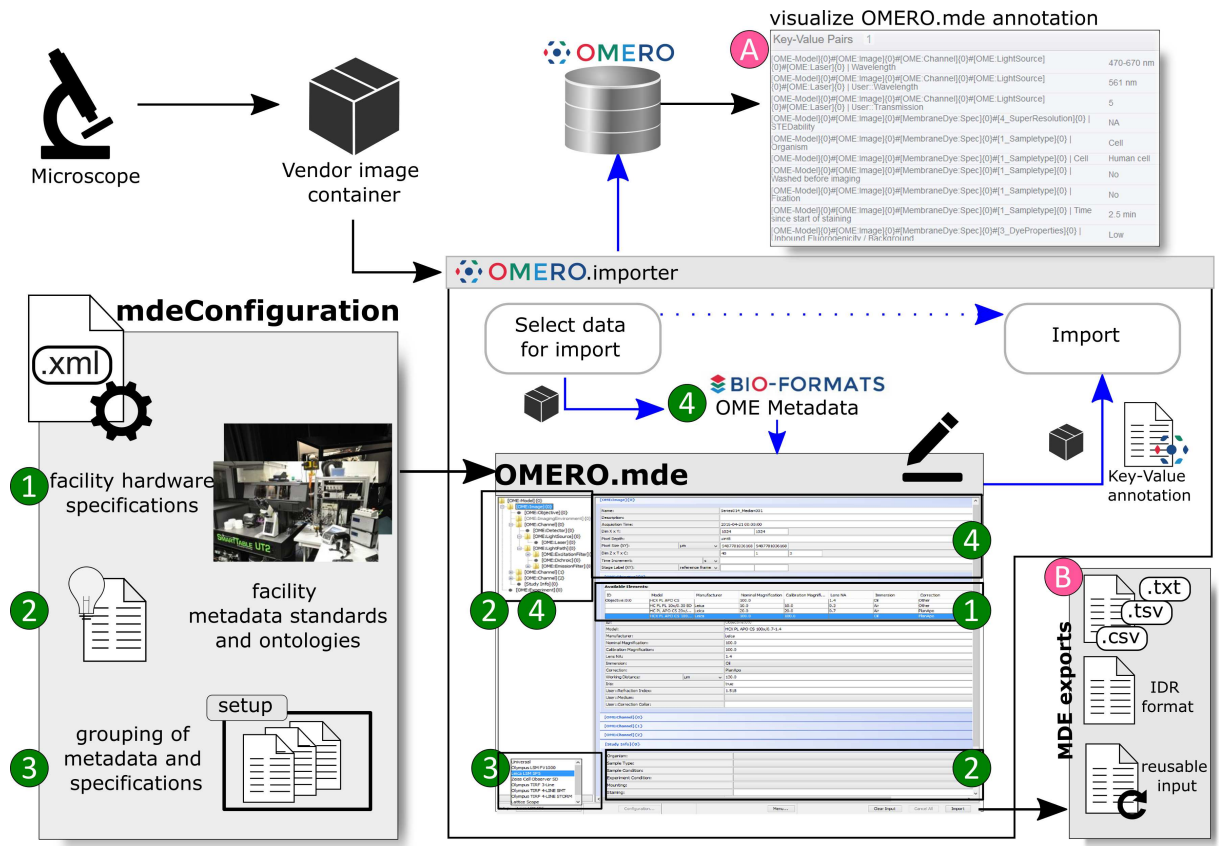}
\caption{\small{\textbf{OMERO.importer with integrated MDEmic as OMERO.mde.} In the OMERO.importer the MDEmic is integrated as an intermediate step for the selection of data for the import and the import itself . Metadata can now be added, which is then transferred to the repository together with the image data (A), or the annotations can be exported in different formats in this step (B). The MDEmic can be customised via a configuration file and loads the specifications from this file dynamically when the OMERO.importer is started (1,2,3). All technical metadata of the images marked in the previous step of data selection are read out by Bio-Format(4) and provided as values in the MDEmic respectively.}}
\label{fig:integration}
\end{figure}

Although today the majority of scientific data including microscopy and imaging data are available in digital format, a real benefit from easy sharing and re-using digital data according to the FAIR principles \cite{Wilkinson2016} only exists if data are understandable and unambiguously interpretable. Collecting and maintaining the relevant metadata is key to ensuring that data are reliable, reusable and can be found and accessed by the scientific community. Imaging data are usually extremely rich data files as they report on various parameters in a multidimensional space and are acquired with complex microscopy instruments. The metadata or data models are very diverse due to the wide range of e.g. modalities, scales, experimental setups and file formats. Therefore, the appropriate use of suitable standardized metadata and data models is a challenge \cite{Linkert2010,Williams2017}. Accordingly, flexible tools for capturing a complete set of metadata are in great demand by researchers applying microscopy techniques. Moreover, it is important for imaging core facilities to be able to provide different standards with one tool and still be flexible enough for dynamic developments. Many tools fail to strike this balance and therefore oscillate between using rigid data models and free text input without semantic context. Similarly, the integration and referencing of existing metadata is often lacking. 
Our tool, MDEmic (MetaData Editor for microscopy), provides an easy and comfortable way for editing metadata of microscopic imaging data and at the same time offers high flexibility in terms of adjustment of metadata sets (Fig.\ref{fig:integration}). As the standardization process regarding the metadata of microscopic experiments is in full swing, MDEmic offers high flexibility to follow this process. This is achieved through the dynamic configurability of both the queried or integrated metadata and predefined values and the reference to ontology databases. MDEmic reads the technical metadata stored in the image file using Bio-Formats \cite{Linkert2010}, a software library for reading proprietary microscopy image (meta)data and presents this metadata in the form of the OME (Open Microscopy Environment) Data Model \cite{Goldberg2005May,OMEFileFormats2020Dec,Swedlow2003}. Visualizing this model as input forms allows the researcher to adjust or correct the technical metadata. In addition, based on the default in a configuration file associated with MDEmic, input forms for further metadata can be generated dynamically, integrating, or extending the OME Data Model for technical metadata. The specification of metadata can include: i) type and category of metadata, ii) fixed terms as selectable input values loading from subclasses by specifying ontology class identifiers, iii) definition of relation to other metadata categories.
\begin{figure}[h]
\centering
\includegraphics[width=1.0\linewidth]{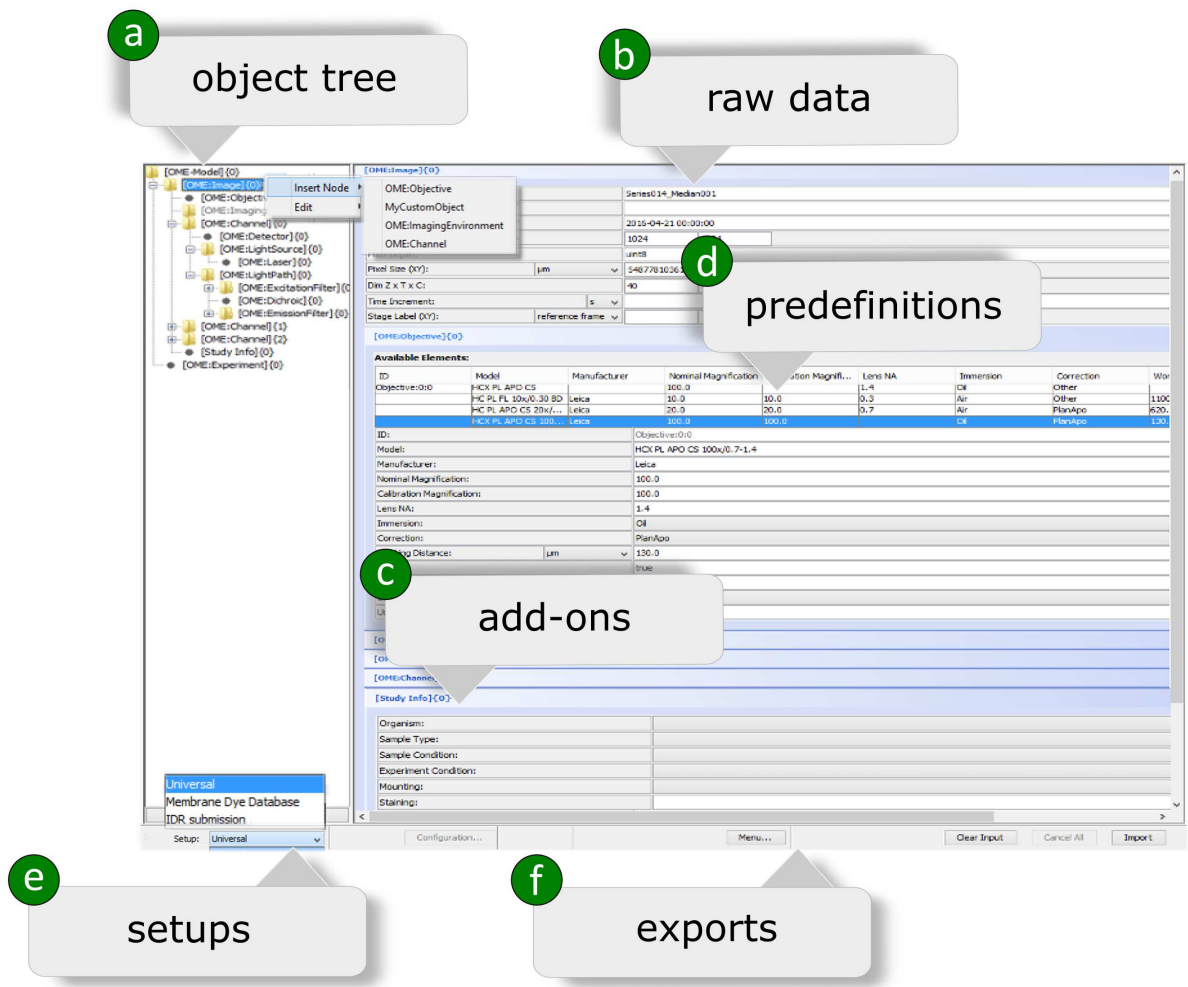}
\caption{\small{\textbf{Graphical user interface of MDEmic.} \textbf{a)}The object tree is based on the structure of the OME data model and can be extended by objects of this data model or self-defined objects. Objects are a collection of related metadata. The extension can be done manually at runtime by using the context menu or by selecting the appropriate setup from the configuration file. \textbf{b)} The technical metadata stored in the image file is read using Bio-Formats and provided in the input forms as editable values. \textbf{c)} The user can define new objects with defined metadata keys and a selection of predefined values by specifying them in the configuration file. These values can also be loaded automatically by reference to an ontology class. \textbf{d)}Predefined metadata can be specified for all objects in the configuration, which the user can choose from. \textbf{e)}A setup is a bundle of data model modifications, input form configurations and/or various associated predefinitions (such as hardware definitions of a microscope setup or an experiment protocol). \textbf{f)}All metadata can be exported directly to a text file or as a reusable template for later annotation using MDEmic.}}
\label{fig:gui}
\end{figure}
\begin{figure}[h]
\centering
\includegraphics[width=1.0\linewidth]{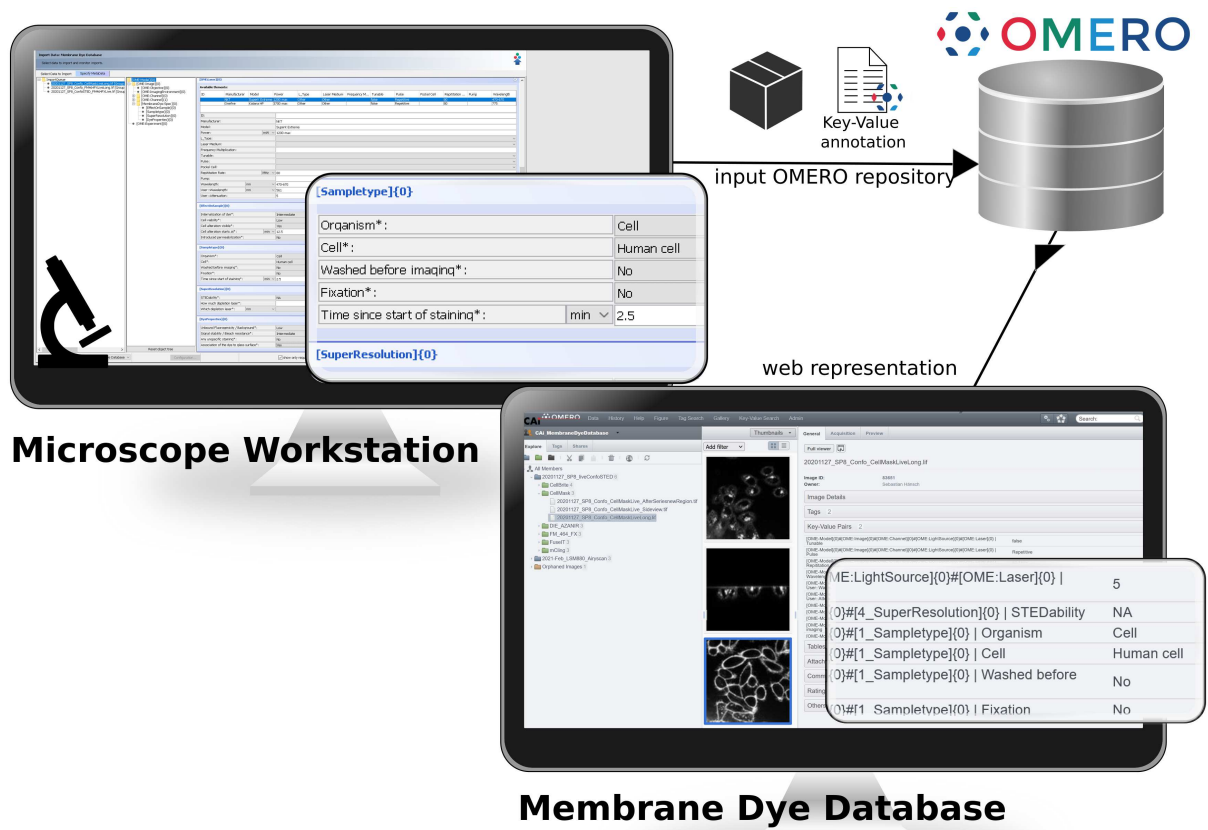}
\caption{\small{\textbf{The workflow from microscopes to (tailored) OMERO repositories.} After data acquisition on the microscope, start the software tool (OMERO.importer) for data transfer to the Membrane Dye Database. For the selected data, the metadata contained in the data container are read out using Bio-Formats. Now the required metadata for a membrane dye dataset can be added using the given input mask. Finally, the data and the added metadata are transferred to the Membrane Dye Database.}}
\label{fig:membrandye}
\end{figure}
\begin{figure}[h]
\centering
\includegraphics[width=1.0\linewidth]{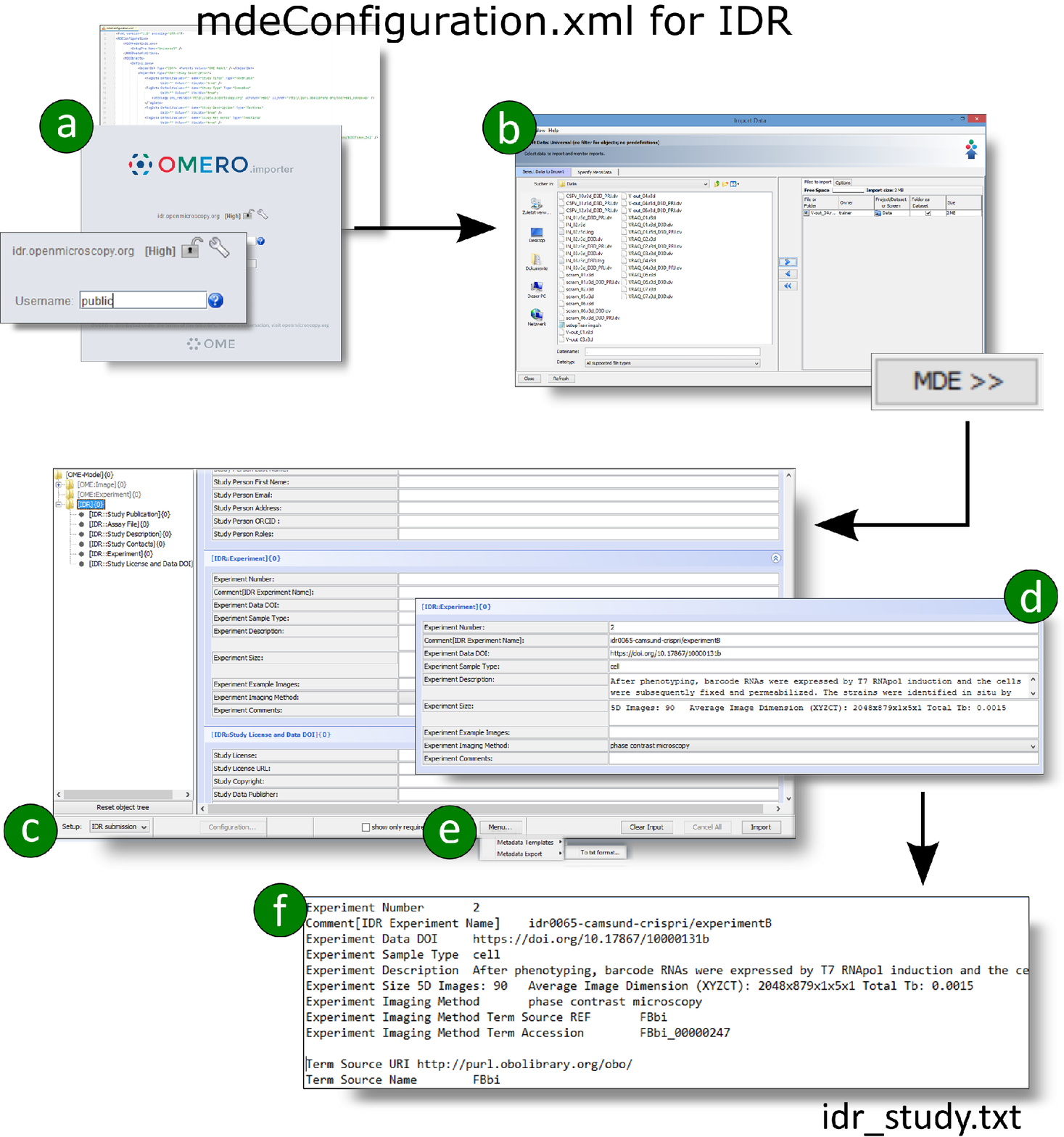}
\caption{\small{\textbf{Capture required metadata for IDR submission with MDEmic integrated in OMERO.} Workflow for collec\-ting the required metadata for an IDR submission using OMERO.mde. \textbf{a)} After downloading the mdeConfiguration.xml (\url{https://doi.org/10.5281/zenodo.5138039}) for the IDR submission , you can adapt the used ontology classes to your needs. OMERO.mde loads all subclass terms from the linked ontology class and makes them available as a selection list. Start the OMERO.importer. By referencing the publicly accessible IDR server, anyone can use the OMERO.importer even without a local OMERO server.  \textbf{b)} Select your data and switch to the mde input field.  \textbf{c)} Select the \textit{IDR submission} setup to load the IDR-specific input masks.  \textbf{d)} Fill in the fields.  \textbf{e)} Exporting the input to an IDR-formatted file automatically creates an  \textbf{f)} idr\_study.txt file with the input as key-value pairs and the references to the selected ontology class.}}
\label{fig:idrWorkflow}
\end{figure}
For all metadata, different sets of predefinitions can be integrated via the configuration file and selected by the user according to the respective scientific application or image technology (Fig.\ref{fig:gui}). MDEmic is part of the standard installation package of the image database OMERO \cite{Allan2012} and is integrated in the OMERO.importer as OMERO.mde. The OMERO.importer can be used without a local existing OMERO installation. All metadata descriptions created in OMERO.mde can be saved and reloaded by the user for later reuse and adaptation or can be exported to different textual formats. This functionality allows the output to be easily integrated with other needs. For example, this increases interoperability with other research data management tools to support integration with other data types or preparation of image data for publications or upload to public repositories \cite{Swedlow2020Oct} like the Image Data Repository, IDR \cite{Williams2017,Ellenberg2018Nov} (IDR workflow Fig.\ref{fig:idrWorkflow}). In the following we describe a use case scenario where image data from samples treated with various membrane dyes are made available in a „Membrane Dye Database“ hosted in the institutional OMERO instance and shared between the members of a collaborative research center (CRC) with a focus in membrane research. Here, we utilized the OMERO.mde for customized metadata annotation. To this purpose we have defined a metadata object „Membrane Dye" in OMERO.mde. This object is available with different sub-objects describing the membrane dyes in more detail like „Effects On Sample" and „Internalization“ of the dye. This object is provided together with the technical description and essential OME Data Model objects as input forms. All input forms are summarized in the setup „Membrane Dye Database" (Fig.\ref{fig:membrandye}). This adjustment for the specific object „Membrane Dye" can be done by editing the configuration file of OMERO.mde e.g., by a data steward of the CRC or a scientist of the imaging core facility. The use case described shows the direct benefit of a metadata annotation tool like MDEmic (in particular here: OMERO.mde) for researchers, as it illustrates a clearly defined purpose. It will help to increase the overall awareness for the importance of metadata annotation and consistent research data management in general.
\subsection*{Acknowledgement}
\paragraph{Funding:}
Supported by Deutsche Forschungsgemeinschaft grants  – SFB 1208 (Projektnummer
267205415 INF/Z02; SFB 944 INF (S.K.); Wellcome Trust grant (Ref: 212962/Z/18/Z) and the BBSRC (Ref: BB/R015384/1) (F.W.); NIH grant (Ref:\\ 5U01CA200059-03), and by Chan Zuckerberg Initiative DAF grant (Ref: 2019-198155 (5022)), an advised fund of Silicon Valley Community Foundation, as part of their Imaging Scientist Program (C.S.D.C).
\paragraph{Author contributions \cite{CRediT}:} S.K.: Software, Writing - Original Draft, Visualization; S.W-P.: Conzeptualization, Writing - Original Draft; S.H.: Visualization, Investigation; F.W.: Writing - Review and Editing; C.S: Writing - Review and Editing; C.S.D.C: Writing - Review and Editing.
\paragraph{Competing Interests:} The authors declare no competing interests.
\paragraph{Data and materials availabilty:}OMERO.mde is implemented in Java, configuration as input and templates as outputs are designed in XML. OMERO as well as OMERO.mde are open source and available under the GNU General public license at \url{https://github.com/ome/omero-insight}. Instructions can be found at
\url{https://omero-guides.readthedocs.io/projects/omero-guide-mde/en/latest/}. The mdeConfiguration.xml for example workflows ("Membran Dye Database" and IDR) are available at \url{https://doi.org/10.5281/zenodo.5138039}.
\begingroup
\raggedright
\sloppy
\printbibliography

@Article{Williams2017,
author={Williams, Eleanor
and Moore, Josh
and Li, Simon W.
and Rustici, Gabriella
and Tarkowska, Aleksandra
and Chessel, Anatole
and Leo, Simone
and Antal, B{\'a}lint
and Ferguson, Richard K.
and Sarkans, Ugis
and Brazma, Alvis
and Carazo Salas, Rafael E.
and Swedlow, Jason R.},
title={Image Data Resource: a bioimage data integration and publication platform},
journal={Nature Methods},
year={2017},
%month={8},
day={01},
volume={14},
number={8},
pages={775-781},
abstract={This Resource describes the Image Data Resource (IDR), a prototype online system for biological image data that links experimental and analytic data across multiple data sets and promotes image data sharing and reanalysis.},
issn={1548-7105},
doi={10.1038/nmeth.4326},
}

@article{Ellenberg2018Nov,
	author = {Ellenberg, Jan and Swedlow, Jason R. and Barlow, Mary and Cook, Charles E. and Sarkans, Ugis and Patwardhan, Ardan and Brazma, Alvis and Birney, Ewan},
	title = {{A call for public archives for biological image data}},
	journal = {Nat. Methods},
	volume = {15},
	pages = {849--854},
	year = {2018},
	%month = {11},
	issn = {1548-7105},
	publisher = {Nature Publishing Group},
	doi = {10.1038/s41592-018-0195-8}
}

@Article{Wilkinson2016,
author={Wilkinson, Mark D.
and Dumontier, Michel
and Aalbersberg, IJsbrand Jan
and Appleton, Gabrielle
and Axton, Myles
and Baak, Arie
and Blomberg, Niklas
and Boiten, Jan-Willem
and da Silva Santos, Luiz Bonino
and Bourne, Philip E.
and Bouwman, Jildau
and Brookes, Anthony J.
and Clark, Tim
and Crosas, Merc{\`e}
and Dillo, Ingrid
and Dumon, Olivier
and Edmunds, Scott
and Evelo, Chris T.
and Finkers, Richard
and Gonzalez-Beltran, Alejandra
and Gray, Alasdair J.G.
and Groth, Paul
and Goble, Carole
and Grethe, Jeffrey S.
and Heringa, Jaap
and 't Hoen, Peter A.C
and Hooft, Rob
and Kuhn, Tobias
and Kok, Ruben
and Kok, Joost
and Lusher, Scott J.
and Martone, Maryann E.
and Mons, Albert
and Packer, Abel L.
and Persson, Bengt
and Rocca-Serra, Philippe
and Roos, Marco
and van Schaik, Rene
and Sansone, Susanna-Assunta
and Schultes, Erik
and Sengstag, Thierry
and Slater, Ted
and Strawn, George
and Swertz, Morris A.
and Thompson, Mark
and van der Lei, Johan
and van Mulligen, Erik
and Velterop, Jan
and Waagmeester, Andra
and Wittenburg, Peter
and Wolstencroft, Katherine
and Zhao, Jun
and Mons, Barend},
title={The FAIR Guiding Principles for scientific data management and stewardship},
journal={Scientific Data},
year={2016},
%month={3},
day={15},
volume={3},
number={1},
pages={160018},
abstract={There is an urgent need to improve the infrastructure supporting the reuse of scholarly data. A diverse set of stakeholders---representing academia, industry, funding agencies, and scholarly publishers---have come together to design and jointly endorse a concise and measureable set of principles that we refer to as the FAIR Data Principles. The intent is that these may act as a guideline for those wishing to enhance the reusability of their data holdings. Distinct from peer initiatives that focus on the human scholar, the FAIR Principles put specific emphasis on enhancing the ability of machines to automatically find and use the data, in addition to supporting its reuse by individuals. This Comment is the first formal publication of the FAIR Principles, and includes the rationale behind them, and some exemplar implementations in the community.},
issn={2052-4463},
doi={10.1038/sdata.2016.18}
}

@Article{Allan2012,
author={Allan, Chris
and Burel, Jean-Marie
and Moore, Josh
and Blackburn, Colin
and Linkert, Melissa
and Loynton, Scott
and MacDonald, Donald
and Moore, William J.
and Neves, Carlos
and Patterson, Andrew
and Porter, Michael
and Tarkowska, Aleksandra
and Loranger, Brian
and Avondo, Jerome
and Lagerstedt, Ingvar
and Lianas, Luca
and Leo, Simone
and Hands, Katherine
and Hay, Ron T.
and Patwardhan, Ardan
and Best, Christoph
and Kleywegt, Gerard J.
and Zanetti, Gianluigi
and Swedlow, Jason R.},
title={OMERO: flexible, model-driven data management for experimental biology},
journal={Nature Methods},
year={2012},
%month={3},
day={01},
volume={9},
number={3},
pages={245-253},
abstract={The Open Microscopy Environment Remote Objects (OMERO) software platform provides a server-based system for managing and analyzing microscopy images and non-image data.},
issn={1548-7105},
doi={10.1038/nmeth.1896},
}

@article{Swedlow2003,
	author = {Swedlow, Jason R. and Goldberg, Ilya and Brauner, Erik and Sorger, Peter K.},
	title = {{Informatics and Quantitative Analysis in Biological Imaging}},
	journal = {Science},
	volume = {300},
	number = {5616},
	pages = {100--102},
	year = {2003},
	%month = {4},
	issn = {0036-8075},
	publisher = {American Association for the Advancement of Science},
	doi = {10.1126/science.1082602}
}

@article{Linkert2010,
	author = {Linkert, Melissa and Rueden, Curtis T. and Allan, Chris and Burel, Jean-Marie and Moore, Will and Patterson, Andrew and Loranger, Brian and Moore, Josh and Neves, Carlos and MacDonald, Donald and Tarkowska, Aleksandra and Sticco, Caitlin and Hill, Emma and Rossner, Mike and Eliceiri, Kevin W. and Swedlow, Jason R.},
	title = {{Metadata matters: access to image data in the real world}},
	journal = {J. Cell Biol.},
	volume = {189},
	number = {5},
	pages = {777--782},
	year = {2010},
	%month = {5},
	issn = {0021-9525},
	publisher = {The Rockefeller University Press},
	doi = {10.1083/jcb.201004104}
}

@article{Goldberg2005May,
	author = {Goldberg, Ilya G. and Allan, Chris and Burel, Jean-Marie and Creager, Doug and Falconi, Andrea and Hochheiser, Harry and Johnston, Josiah and Mellen, Jeff and Sorger, Peter K. and Swedlow, Jason R.},
	title = {{The Open Microscopy Environment (OME) Data Model and XML file: open tools for informatics and quantitative analysis in biological imaging}},
	journal = {Genome Biol.},
	volume = {6},
	number = {5},
	pages = {1--13},
	year = {2005},
	%month = {5},
	issn = {1474-760X},
	publisher = {BioMed Central},
	doi = {10.1186/gb-2005-6-5-r47}
}

@misc{OMEFileFormats2020Dec,
	title = {{OME Data Model and File Formats 6.2.2 Documentation {\ifmmode---\else\textemdash\fi} OME Data Model and File Formats 6.2.2 documentation}},
	year = {2020},
	%month = {12},
	%note = {[Online; accessed 24. Feb. 2021]},
	note = {url: \url{https://docs.openmicroscopy.org/}\\\url{ome-model/6.2.2}}
}

@article{Swedlow2020Oct,
	author = {Swedlow, Jason R. and Kankaanp{\ifmmode\ddot{a}\else\"{a}\fi}{\ifmmode\ddot{a}\else\"{a}\fi}, Pasi and Sarkans, Ugis and Goscinski, Wojtek and Galloway, Graham and Sullivan, Ryan P. and Brown, Claire M. and Wood, Chris and Keppler, Antje and Loos, Ben and Zullino, Sara and Longo, Dario Livio and Aime, Silvio and Onami, Shuichi},
	title = {{A Global View of Standards for Open Image Data Formats and Repositories}},
	journal = {arXiv},
	year = {2020},
	%month = {10},
	note = {url: \url{https://arxiv.org/abs/2010.10107v1}}
}

@misc{CRediT,
	author = {Elsevier},
	title = {{CRediT author statement}},
	year = {2021},
	%month = {2},
	%note = {[Online; accessed 24. Feb. 2021]},
	note={url: \url{https://www.elsevier.com/authors/}\\\url{policies-and-guidelines/credit-author-statement}}
	%url = {https://www.elsevier.com/authors/policies-and-guidelines/credit-author-statement}
}
\endgroup

\end{document}